\documentclass[pra,twocolumn]{revtex4-2}
\usepackage{eurosym}
\usepackage{amssymb}
\usepackage{amsfonts}
\usepackage{amsmath}
\usepackage{graphicx}
\usepackage{bm}
\usepackage{braket}
\usepackage{epstopdf}
\usepackage{xcolor}

\begin{document}

\title{Bulk modulus of three-dimensional quantum droplets}

\author{Zibin Zhao$^{1}$}
\author{Guilong Li$^{2}$}
\author{Zhaopin Chen$^{3}$}
\author{Huan-Bo Luo$^{1,4}$}
\author{Bin Liu$^{1,4}$}
\email{binliu@fosu.edu.cn}
\author{Boris A. Malomed$^{5,6}$}
\author{Yongyao Li$^{1,4}$ }
\email{yongyaoli@gmail.com}
\affiliation{$^1$School of Physics and Optoelectronic Engineering, Foshan University,
Foshan 528000, China\\
$^2$College of Engineering and Applied Sciences, National Laboratory of Solid State
Microstructures, Nanjing University, Nanjing 210023, China\\
$^3$Physics Department and Solid-State Institute, Technion, Haifa 32000, Israel\\
$^{4}$Guangdong-Hong Kong-Macao Joint Laboratory for Intelligent Micro-Nano
Optoelectronic Technology, Foshan University, Foshan 528225, China\\
$^5$Department of Physical Electronics, School of Electrical Engineering,
Faculty of Engineering, Tel Aviv University, Tel Aviv 69978, Israel\\
$^6$Instituto de Alta Investigaci\'{o}n, Universidad de Tarapac\'{a},
Casilla 7D, Arica, Chile }

\begin{abstract}
Quantum droplets (QDs), formed by ultra-dilute quantum fluids under the action of
Lee-Huang-Yang (LHY) effect, provide a unique platform for investigating a
wide range of macroscopic quantum effects. Recent studies of QDs' breathing
modes and collisional dynamics have revealed their compressibility and
extensibility, which suggests that their elasticity parameters can be identified. In
this work, we derive the elastic bulk modulus (BM) of QDs by means of the
theoretical analysis and numerical simulations, and establish a relation
between the BM and the eigenfrequency of QD's intrinsic vibrations. The
analysis reveals the dependence of the QD's elasticity on the particle number and
the strength of interparticle interactions. We additionally provide
a realistic estimate of the bulk modulus for the system, yielding a concrete physical
value that may serve as a reference for future experimental measurements. Taken together, 
these results also point to possibilities for realizing elastic media governed by the LHY effect.
\end{abstract}

\maketitle

\section{Introduction}

In classical continua, elasticity characterizes the ability of a material to
resist deformation and store mechanical energy \cite%
{lifshitz-chart-one,Reddy-continuum}. The fundamental elasticity coefficient
is the bulk modulus (BM) $B$, which quantifies the resistance to isotropic
compression, being the single elastic parameter in ordinary liquids, which
lack shear rigidity. The BM governs compressional modes, such as breathing
oscillations, whose frequency is directly determined by the pressure
response to density perturbations.

In the framework of ultracold quantum gases, the balance between the
competing intrinsic nonlinearities, \textit{viz}., mean-field (MF)
attraction and beyond-MF (Lee-Huang-Yang, LHY) repulsion, stabilizes quantum
droplets (QDs), i.e., self-bound states of the ultradilute quantum fluid~%
\cite{LHY-term,Petrov2015,Petrov-low-dimensional,SB_qds,scien-qds}. The superfluid density in
QDs is extremely low, yet giving rise to the effective incompressibility and
macroscopic behavior reminiscent of ordinary liquids, provided that the size
of these states is much larger than the thickness of the \textquotedblleft
skin layer" at their surface (therefore they are named \textquotedblleft
droplets", assuming that the surface energy is small vs. the bulk energy).
In conventional fluids, the BM originates from short-range interatomic
interactions, whereas in QDs it emerges predominantly from the interplay
of contact interaction and quantum fluctuations~\cite{LUO-FOP-rev,TPfau-rev,
bottcherNewstates2020}. Examining
the BM thus provides direct insight into how the quantum pressure, arising
from the balance of the MF and beyond-MF interactions, responds to changes
in the density or interaction strength. This offers a quantitative means to
characterize the QD fluidity and converts the LHY effect of quantum
fluctuations into an experimentally accessible quantity.

In the experiment, the precise control of interatomic interactions~is
realized by means of the Feshbach resonances \cite{FRRu87,FRRMP,OFRPRL},
allowing one to manipulate the system's parameters and excite intrinsic
collective modes \cite{pollack-collective, shi-23NA, stein-photon, XCui-QFDD, hhu-pra,
petrov-collective, Santos-impurities-1Dqds, YFei-2D-qds, Blakie_collective, DynaformK39,
PRL-multipld-QDs,XDu-Bogo-mode-1Dqds,Englezos-correlat-dyna,hhu_CE_sphere}, such as breathing
oscillations \cite{Guebli-HG-quam-fluc, sturmer-breath-mode, LZLv, Rajat-collective-spin1,
Bougas-2D, orignac-breathing-mode-dopolar, Boris_1Ds} and higher-order
vibrational modes \cite{Pathak-SR, Wysocki-Josephson-dynamics}, by means of
the interaction quench \cite{YNie-spectra-OL, homo-tereo, Gangwar-spc-QDs,
Pelayo-fermi-impurities}, collision \cite{XHu-scattering-1Dqds, Collisions_QDs,
DECqds, Collision_2Dqds, AYang-dipolar,Salgueiro-vortex-QDs, Li-Elongated-PRA,
Li-dipolar-QDs}, or similar protocols
\cite{Dauer-floquet resonances, Maity-Parametrically excited, clark-collective,
Wu-matter-wave jet, DLW-rotating}. However, to date, studies of the
QD dynamics were focused on breathing oscillations and other
intrinsic collective modes, while the underlying elastic modulus,
that governs the dynamical behavior and determines
eigenfrequencies of the breathing modes, was not addressed.

In this work, by employing analytical and numerical methods, we determine
the QD's breathing-mode eigenfrequency $\Omega $ and BM $B$, elucidating
their dependence on the strength of the interatomic interactions and
particle number. By defining ratio $\eta \equiv B/\Omega ^{2}$, we establish
a direct connection between $B$ and $\Omega $, with their ratio $\eta $
determined by the total atom number and strength of the interaction between atoms.
The remainder of this paper is organized as follows. Section~\ref{sec:model}
introduces the theoretical model. Section~\ref{sec:VA} presents the variational
approximation for the ground state. Section~\ref{sec:comparison} compares the
numerical results with the predictions of the variational analysis.
Section~\ref{sec:relation} discusses the relation between the bulk modulus $B$
and the intrinsic excitation frequency $\Omega$. Section~\ref{sec:experiment}
provides experimental estimates by converting the dimensionless quantities into
their corresponding physical values. Section~\ref{sec:conclusion} concludes the
paper.

\section{The Model}
\label{sec:model}
We consider a binary BEC of $^{39}\mathrm{K}$ atoms
in three-dimensional space with coordinates $(X,Y,Z)$, modeled by a system
of nonlinearly coupled Gross--Pitaevskii equations that include both the
cubic mean-field terms and the quartic Lee--Huang--Yang (LHY)
corrections~\cite{Petrov2015}:

\begin{equation}
    \begin{split}
i\hbar \frac{\partial }{\partial T}\Psi _{1}=&-\frac{\hbar ^{2}}{2M}\nabla _{%
\mathrm{XYZ}}^{2}\Psi _{1}
+\left( G_{11}|\Psi _{1}|^{2}+G_{12}|\Psi
_{2}|^{2}\right) \Psi _{1}\\
&+\Upsilon \left( |\Psi _{1}|^{2}+|\Psi
_{2}|^{2}\right) ^{\frac{3}{2}}\Psi _{1},  \label{coupled-GPE1}
    \end{split}
\end{equation}%
\begin{equation}
    \begin{split}
        i\hbar \frac{\partial }{\partial T}\Psi _{2}=&-\frac{\hbar ^{2}}{2M}\nabla _{%
        \mathrm{XYZ}}^{2}\Psi _{2}+\left( G_{22}|\Psi _{2}|^{2}+G_{21}|\Psi
        _{1}|^{2}\right) \Psi _{2}\\
        &+\Upsilon \left( |\Psi _{1}|^{2}+|\Psi
        _{2}|^{2}\right) ^{\frac{3}{2}}\Psi _{2},  \label{coupled-GPE2}
    \end{split}
\end{equation}%
where $G_{11}=G_{22}=4\pi \hbar^{2}a/M$ and $G_{12}=G_{21}=4\pi \hbar^{2}a'/M$
are the self- and cross-interaction strengths, with atomic mass $M$, and
$a$ and $a'$ the intra- and inter-species scattering lengths, respectively,
while $\Psi_{1}$ and $\Psi_{2}$ represent the macroscopic wave functions of the
two components. The coefficient of the LHY correction is
\cite{Petrov2015}%
\begin{equation}
\Upsilon =\frac{128\sqrt{\pi }}{3M}\hbar ^{2}a^{\frac{5}{2}}.  \label{LHY}
\end{equation}%
For symmetric states in the binary BEC, with
\begin{equation}
\Psi _{1}=\Psi _{2}\equiv \Psi /\sqrt{2},  \label{symmetric-condition}
\end{equation}%
Eqs. (\ref{coupled-GPE1}) and (\ref{coupled-GPE2}) admit the reduction to a
single equation,
\begin{equation}
i\hbar \frac{\partial }{\partial T}\Psi =-\frac{\hbar ^{2}}{2M}\nabla _{%
\mathrm{XYZ}}^{2}\Psi +\frac{\delta G}{2}\left\vert \Psi \right\vert
^{2}\Psi +\Upsilon \left\vert \Psi \right\vert ^{3}\Psi ,  \label{GPE}
\end{equation}%
where $\delta G=\left( 4\pi \hbar ^{2}/M\right) \left( a^{\prime }+a\right)
\equiv \left( 4\pi \hbar ^{2}/M\right) \delta a$, and $\delta a=a^{\prime
}+a $.
The total number of atoms in the system is
\begin{equation}
N=\int \left( \left\vert \Psi _{1}\right\vert ^{2}+\left\vert \Psi
_{2}\right\vert ^{2}\right) d^{3}\mathbf{R}=\int \left\vert \Psi \right\vert
^{2}d^{3}\mathbf{R}.  \label{atom-Number}
\end{equation}%
By means of rescaling,
\begin{equation}
T\equiv t_{0}t,\left( X,Y,Z\right) \equiv l_{0}\left( x,y,z\right) ,\Psi
\equiv l_{0}^{-\frac{3}{2}}\psi ,  \label{rescaling}
\end{equation}%
where $t_{0}\equiv Ml_{0}^{2}/\hbar $ and $l_{0}$ are time and length
scales, Eq. (\ref{GPE}) is cast in the dimensionless form:
\begin{equation}
i\frac{\partial }{\partial t}\psi =-\frac{1}{2}\nabla ^{2}\psi +g\left\vert
\psi \right\vert ^{2}\psi +\gamma \left\vert \psi \right\vert ^{3}\psi .
\label{dimensionless-GPE-with-gamma}
\end{equation}%
Here, $g=2\pi \delta a/l_{0}<0$ denotes the dimensionless strength of the
effective contact attraction, and $\gamma =\left( 128\sqrt{\pi }/3\right)
\left( a/l_{0}\right) ^{\frac{5}{2}}>0$ represents the dimensionless LHY
correction. In this work, we fix the intra-species scattering length as %
$a=50a_{0}$ ($a_{0}$ is the Bohr radius) and set $l_0=0.1\mu\mathrm{m} $,%
which lies within the experimentally relevant range for QDs,
while being expedient for simulations of the dimensionless systems %
\cite{SB_qds,scien-qds}.
Then, we further rescale Eq.~(\ref%
{dimensionless-GPE-with-gamma}) by setting
\begin{equation*}
t=t\gamma ,\qquad (x,y,z)=(x,y,z)\sqrt{\gamma },\qquad g=g/\gamma ,
\end{equation*}%
which leads to the scaled form
\begin{equation}
i\frac{\partial \psi }{\partial t}=-\frac{1}{2}\nabla ^{2}\psi +g|\psi
|^{2}\psi +|\psi |^{3}\psi ,  \label{dimensionless-GPE}
\end{equation}%
where $g<0$ is the reduced contact-interaction strength. The total norm,
\begin{equation}
\mathcal{N}=\int |\psi |^{2}\,d^{3}\mathbf{r},  \label{norm}
\end{equation}%
is proportional to the total number of atoms in the system.
The Hamiltonian (energy) corresponding to Eq. (\ref{dimensionless-GPE}) is
\begin{equation}
E=\int \Big(\frac{1}{2}\left\vert \nabla \psi \right\vert ^{2}+\frac{1}{2}%
g\left\vert \psi \right\vert ^{4}+\frac{2}{5}\left\vert \psi \right\vert ^{5}%
\Big)d^{3}\mathbf{r}.  \label{energy}
\end{equation}%
Stationary states are look for in the usual form, $\psi (\mathbf{r},t)=\phi (%
\mathbf{r})e^{-i\mu t}$, where $\mu $ is a real chemical potential, and $%
\phi $ the stationary wave function which obeys the spatial GPE:
\begin{equation}
\mu \phi =-\frac{1}{2}\nabla ^{2}\phi +g\left\vert \phi \right\vert ^{2}\phi
+\left\vert \phi \right\vert ^{3}\phi .  \label{stationary-GPE}
\end{equation}
In the framework of the variational
approximation (VA), the chemical potential of a self-bound droplet
in equilibrium case can be calculated as
\begin{equation}
\mu = d E_\mathrm{eq}/ d \mathcal{N},  \label{muVA}
\end{equation}
where $E_\mathrm{eq}$ is the energy evaluated with the equilibrium volume,
determined by the zero-pressure condition
$p=-\left(\partial E/\partial V\right)_{\mathcal{N}}=0$.
For the ground state QD with a fixed atoms number and approximately
homogeneous density, the elastic BM is defined as \cite{BEC-Ol}
\begin{equation}
B=-V\frac{\partial p}{\partial V}=-\mathcal{N}\frac{\partial \mu }{\partial V%
},  \label{Bulkmodulus}
\end{equation}%
where $p=-\partial E/\partial V$ is the pressure, $E$ is the total energy %
(\ref{energy}), $V$ is the QD's volume, and $\mathcal{N}$ is the norm (\ref%
{norm}) (see Appendix \ref{derivation-B} for the derivation of Eq.~(\ref%
{Bulkmodulus})).

\section{The Variational Approximation For The ground states}
\label{sec:VA}
The Lagrangian density corresponding to Eq. (\ref{dimensionless-GPE}) for
the isotropic ground-state (GD) QD, with $\psi (\mathbf{r},t)=\psi \left(
r\equiv \sqrt{x^{2}+y^{2}+z^{2}},t\right) $, is
\begin{equation}
\mathcal{L}=\frac{i}{2}\left( \psi ^{\ast }\partial _{t}\psi -\psi \partial
_{t}\psi ^{\ast }\right) -\frac{1}{2}|\partial _{r}\psi |^{2}-\frac{1}{2}%
g|\psi |^{4}-\frac{2}{5}|\psi |^{5}.  \label{Lagrangian-density}
\end{equation}%
The full Lagrangian being
\begin{equation}
L=4\pi \int \mathcal{L}r^{2}dr.  \label{L}
\end{equation}%
The VA for the QDs is chosen as the super-Gaussian of order $%
\alpha $, which is relevant for models with competing nonlinearities, such
as the previously studied cubic-quintic combinations \cite%
{super1,super2,VA-flat-top,SG-3Ds,VA_2Dqds},
\begin{equation}
\psi (r,t)=A(t)\exp \left\{ -\frac{1}{2}\left[ \frac{r}{w(t)}\right] ^{2\alpha
}+i\beta(t) r^{2} \right\} ,  \label{super-Gaussian}
\end{equation}%
where variational parameters $A(t),w(t)$, and $\beta (t)$
represent the amplitude, width, and chirp, respectively. In particular,
the argument of the complex amplitude $A$ represents the overall phase
of ansatz (\ref{super-Gaussian}).  The
condition of the absence of the singularity at $r=0$, produced by the
substitution of \textit{ansatz} (\ref{super-Gaussian}) in Eq. (\ref%
{dimensionless-GPE}) imposes the constraint $\alpha \geq 1$.

For the GS stationary solution, we set $\beta =0$. Note that
the amplitude $A$ can be determined by the normalization condition,
\begin{equation}
\mathcal{N}=\int d^{3}\mathbf{r}|\psi (\mathbf{r})|^{2}=\frac{2\pi }{\alpha }%
\Gamma _{3}A^{2}w^{3},  \label{A}
\end{equation}%
where $\Gamma _{q}\equiv \Gamma\!\left( q/2\alpha \right) $ is the Gamma
function. For the nonstationary isotropic QD, two
remaining degrees of freedom of the VA \textit{ansatz} are $w(t)$ and $\beta
(t)$. An additional variational parameter is $\alpha $, which we consider as
a time-independent constant. According to \textit{ansatz} (\ref%
{super-Gaussian}), the mean QD's radius is defined by $w$ and $\alpha $ as
\begin{equation}
\bar{r}=\sqrt{\frac{\int r^{2}|\psi |^{2}d^{3}\mathbf{r}}{\int |\psi
|^{2}d^{3}\mathbf{r}}}=w\sqrt{\frac{\Gamma _{5}}{\Gamma _{3}}}.
\label{mean-r}
\end{equation}%
In terms of the mean radius, the QD's volume is
\begin{equation}
V=\frac{4\pi }{3}\bar{r}^{3}.  \label{V}
\end{equation}

The substitution of the super-Gaussian \textit{ansatz} (\ref%
{super-Gaussian}) in Lagrangian (\ref{L}) yields the following Lagrangian
\begin{equation}
    \begin{split}
        L=& -{\mathcal{N}\dot{\beta}w^2}C_{11}-\frac{%
\alpha ^{2}\mathcal{N}}{w^{2}}C_{12}
-2{\mathcal{N}\beta^2 w^2}C_{11}\\ &-%
g\frac{\alpha \mathcal{N}^{2}}{w^{3}}%
C_{13}-\left( %
\frac{\alpha ^{3}\mathcal{N}^{5}}{w^{9}}\right) ^{\frac{1}{2}}%
C_{14}
\left( \frac{2}{5}\right) ^{\frac{3}{2\alpha }},
\label{L-beta-w}
    \end{split}
\end{equation}
with coefficients $C_{ij}$ given by
\begin{equation}
    \begin{split}
       & C_{11}=\frac{\Gamma _{5}}{\Gamma _{3}},
          C_{12}=\frac{\Gamma _{4\alpha +1}}{2\Gamma _{3}},\\
        &C_{13}=\dfrac{1}{4\pi\cdot 2^{3/2\alpha }\Gamma _{3}}, %
        C_{14}=\dfrac{1}{5\sqrt{2}\left( \pi \Gamma
_{3}\right) ^{3/2}}\left( \dfrac{2}{5}\right) ^{\frac{3}{2\alpha }}.
    \end{split}
\end{equation}
For chirp $\beta$, the Euler-Lagrangian equation yields:
\begin{equation}
    \beta = \frac{\dot{w}}{2w}.\label{Eoula-Eqs1}
\end{equation}
Substituting expression (\ref{Eoula-Eqs1}) in the Lagrangian (\ref{L-beta-w})
eliminates $\beta$ and leads to an effective Lagrangian that depends solely on
the width $w$:
\begin{equation}
    \begin{split}
        L=& -\mathcal{N}\ddot{w}w\frac{C_{11}}{2}-\frac{%
\alpha ^{2}\mathcal{N}}{w^{2}}C_{12}\\ &-%
g\frac{\alpha \mathcal{N}^{2}}{w^{3}}%
C_{13}-\left( %
\frac{\alpha ^{3}\mathcal{N}^{5}}{w^{9}}\right) ^{\frac{1}{2}}%
C_{14}
\left( \frac{2}{5}\right) ^{\frac{3}{2\alpha }}.
\label{L-w}
    \end{split}
\end{equation}
The VA produces the following Euler-Lagrangian equations
of motion for $w$:
\begin{eqnarray}
\ddot{w} &=&C_{21}\frac{\alpha ^{2}}{w^{3}}+gC_{22}\frac{\mathcal{N}\alpha }{%
w^{4}}+w^{1/2}C_{23}\left( \frac{\mathcal{N}\alpha }{w^{4}}\right) ^{3/2}
\notag \\
&\mathbf{\equiv }&-\frac{dU}{dw},  \label{Eoula-Eqs2}
\end{eqnarray}%
with coefficients $C_{ij}$ given by
\begin{equation}
\begin{split}
& C_{21}=\frac{\Gamma _{4\alpha +1}}{\Gamma _{5}},C_{22}=\frac{3}{(4\pi
)\cdot 2^{3/2\alpha }\Gamma _{5}},\\
& C_{23}=\frac{9}{10\sqrt{2}\left( \pi
^{3}\Gamma _{3}\right) ^{1/2}\Gamma _{5}}\left( \dfrac{2}{5}\right) ^{\frac{3%
}{2\alpha }}.
\end{split}
\label{C-matrix-123}
\end{equation}

Eq.~(\ref{Eoula-Eqs1}) indicates that chirp $\beta (t)$ can be
expressed, as usual, in terms of $w(t)$, and the effective potential is
\begin{equation}
U(w)=\frac{C_{21}}{2}\frac{\alpha ^{2}}{w^{2}}+\frac{C_{22}}{3}g\frac{%
\mathcal{N}\alpha }{w^{3}}+\frac{2C_{23}}{9}\left(\frac{ \mathcal{N}\alpha
}{w^{3}}\right)^{3/2}.
\end{equation}%
The variation of the oscillating width can be defined as $%
w(t)=w_{0}+\delta w$, where $w_{0}$ is the equilibrium value, and $\delta w$
represents a small deviation from it. Substituting this expression in Eq.~(%
\ref{Eoula-Eqs2}) and linearizing it with respect to $\delta w$, we obtain
\begin{equation}
\frac{d^{2}}{dt^{2}}\delta w+\Omega ^{2}\,\delta w=0,  \label{d-omega-eq}
\end{equation}%
where
\begin{equation}
\Omega _{\mathrm{VA}}^{2}=\frac{d^{2}U}{dw^{2}}=3C_{21}\frac{\alpha ^{2}}{%
w_{0}^{4}}+4gC_{22}\frac{\mathcal{N}\alpha }{w_{0}^{5}}+\frac{11}{2}C_{23}w_{0}\left(
\frac{\mathcal{N}\alpha }{w_{0}^{5}}\right) ^{3/2}  \label{Omega2}
\end{equation}
is the squared eigenfrequency of the intrinsic vibrations. Note that we
employ a small-perturbation expansion, with $|\delta w| \ll w_0$.
In this regime, the oscillations of the droplet width $w$ are treated as
first-order fluctuations that preserve the fundamental profile. Therefore,
the assumption of the time-invariant $\alpha$ is well-founded. This approach
is consistent with the methodologies adopted in Refs.~\cite{SG-3Ds,VA_2Dqds}.

\begin{figure}[tbp]
{\includegraphics[width=3.4in]{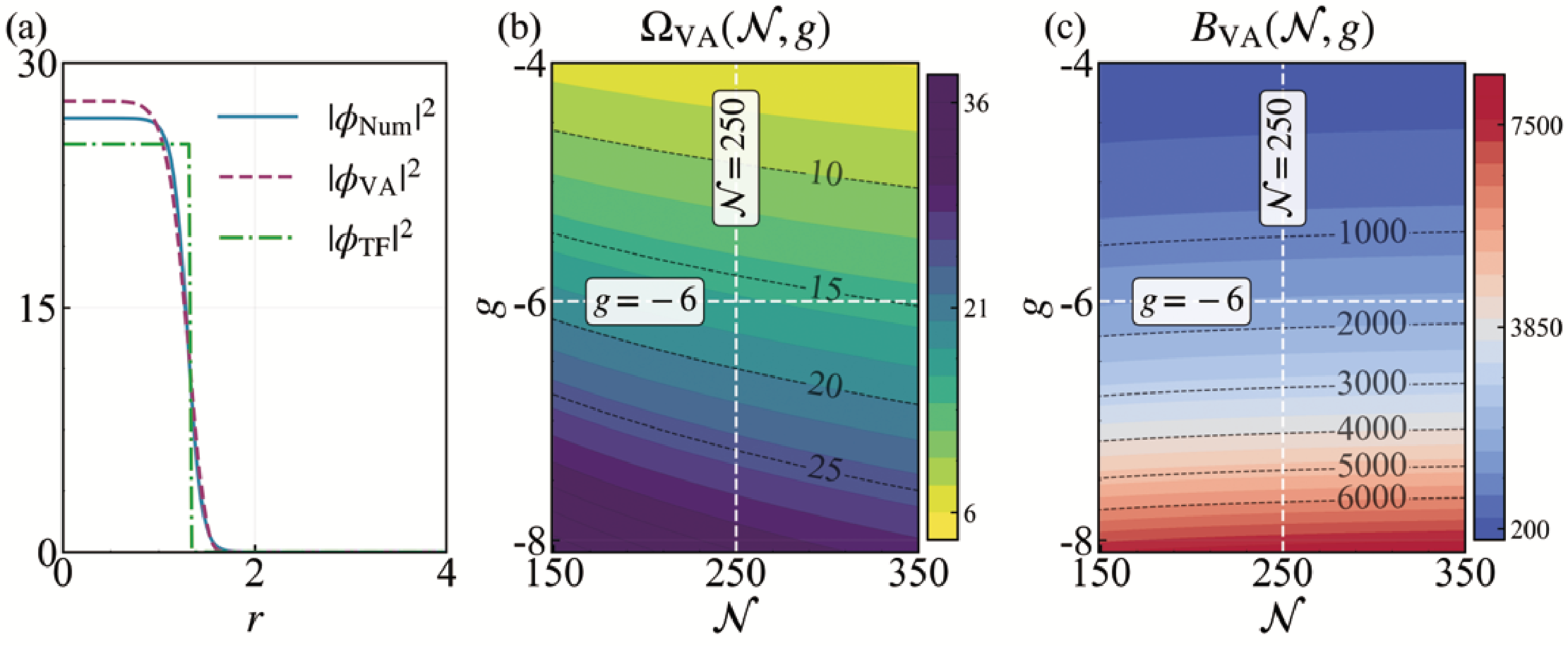}}
\caption{(a) The radial density distribution of the stationary isotropic QD
for $\mathcal{N}=250$ and $g=-6$. The blue solid curve represents the
numerical result, the purple dashed curve shows the VA prediction,
obtained with the super-Gaussian\textit{\ ansatz}, and the green
dash-dotted curve is the TF result.} (b,c) Heatmaps of the
VA-predicted values of the eigenfrequency of the internal oscillations $%
\Omega (\mathcal{N},g)$ and BM $B(\mathcal{N},g)$ [see Eqs. (\protect\ref%
{Omega2}) and (\protect\ref{VA-Bulkmodulus})], in the plane of norm $%
\mathcal{N}$ and reduced MF interaction strength $g$. The color shading from
light to dark indicates increasing values of $\Omega $ and $B$, with the
black dashed curves representing their contour lines. The two white dashed
lines correspond to $\mathcal{N}=250$ and $g=-6$, which represent the cases
shown in Figs. \protect\ref{B-omega-gN} (a,c) and (b,d), respectively.
\label{pk-omega-B}
\end{figure}

For the steady-state solution, the equilibrium value of $w$ can be obtained
from Eq. (\ref{Eoula-Eqs2}), setting $\ddot{w}=0$ in it, while the
steady-state value of the super-Gaussian order $\alpha $ originates from the
variational condition $\partial L/\partial \alpha =0$. These conditions
yield a system of coupled algebraic equations for $\alpha $ and $w$. Solving
it numerically, the stationary values of $w$ and $\alpha $ can be obtained.
In Fig.~\ref{pk-omega-B}(a), we compare the radial density distributions of
the numerically found $\left( \phi _{\mathrm{Num}}\right) $, VA-predicted
$\left( \phi _{\mathrm{VA}}\right) $  and Thomas-Fermi (TF)
$\left( \phi _{\mathrm{TF}}\right) $ solutions for the
stationary isotropic
QD with $\mathcal{N}=250$ and $g=-6$, the respective VA parameters being $%
w=1.34$ and $\alpha =4.57$ [note that this value of $\alpha $ is large in
comparison to $\alpha =1$, which corresponds to the usual Gaussian, see Eq. (%
\ref{super-Gaussian})]. It is seen that the numerical and VA profiles almost
coincide, corroborating the accuracy of the VA based on the super-Gaussian \textit{%
ansatz}. Furthermore, compared to the TF approximation, the
VA provides a significantly better description of the decay characteristics
in the distribution tails. Note that both the numerically exact and approximate radial profiles
plotted in Fig.~\ref{pk-omega-B}(a) satisfy the above-mentioned condition, that
its overall size must be much larger than the thickness of the surface layer.

Further, the VA expression for energy, chemical potential, and BM can be derived
from Eqs. (\ref{super-Gaussian}), (\ref{energy}), (\ref{muVA}), and (\ref{Bulkmodulus}) as
\begin{equation}
E_{\mathrm{VA}}=\mathcal{N}\left[ C_{12}\frac{\alpha ^{2}}{w^{2}}+gC_{13}%
\frac{\mathcal{N}\alpha }{w^{3}}+C_{14}\left( \frac{\mathcal{N}\alpha }{w^{3}%
}\right) ^{3/2}\right] ,  \label{energy-VA}
\end{equation}
\begin{equation}
\mu _{\mathrm{VA}}=C_{12}\frac{%
\alpha ^{2}}{w^{2}}+2gC_{13}\frac{\mathcal{N}\alpha }{w^{3}}+\frac{5}{2}%
C_{14}\left( \frac{\mathcal{N}\alpha }{w^{3}}\right) ^{3/2},  \label{mu}
\end{equation}%
\begin{equation}
\begin{split}
B_{\mathrm{VA}}
=\frac{\mathcal{N}}{V}\left[ \frac{2}{3}C_{12}\frac{\alpha ^{2}}{w^{2}}%
+2gC_{13}\frac{\mathcal{N}\alpha }{w^{3}}+\frac{15}{4}C_{14}\left( \frac{\mathcal{N}\alpha }{%
w^{3}}\right) ^{3/2}\right].
\end{split}
\label{VA-Bulkmodulus}
\end{equation}%

According to Eqs.~(\ref{Omega2}) and (\ref{VA-Bulkmodulus}), we plot the
VA-predicted eigenfrequency $\Omega (\mathcal{N},g)$ of the internal
excitations and BM $B(\mathcal{N},g)$ in Figs.~\ref{pk-omega-B}(b,c). It is
seen that, as the total norm $\mathcal{N}$ increases, $\Omega $ gradually
decreases, while $B$ increases. In contrast, as $-g$
increases---corresponding to a stronger MF\ attraction---both $\Omega $ and $%
B$ increase.

\section{COMPARISON OF NUMERICAL AND VA RESULTS}
\label{sec:comparison}
To extract values the vibration frequency $\Omega _{\mathrm{Num}}$ and BM $%
B_{\mathrm{Num}}$ of the QD from the numerical simulations, we applied
quench of the interaction strength $g$ and analyzed the subsequent dynamics.
Specifically, we monitored the evolution of the chemical potential $\mu (t)$
and effective volume $V(t)$, from which $\Omega _{\mathrm{Num}}$ and $B_{%
\mathrm{Num}}$ can be obtained [in the latter case, by means of Eq. (\ref%
{Bulkmodulus})]. The detailed procedure is as follows:

\begin{enumerate}
\item \textbf{The GS preparation:} For given parameters $(\mathcal{N},g)$,
we solved Eq.~(\ref{dimensionless-GPE}), using the imaginary-time method %
\cite{yangNonlinear2010} to produce a stable GS QD.

\item \textbf{The application of the interaction quench and subsequent
real-time evolution:} The interaction strength was suddenly changed, $%
g\rightarrow g+\delta g$, with $\delta g/g=0.01$. The system was then
evolving in real time, while the time-dependent chemical potential
$\mu (t)=\int \psi^*\hat{H}\psi d^3\mathrm{r}/\mathcal{N}$
and effective volume $V(t)$ were recorded.

\item \textbf{The} \textbf{determination of the BM:} The trajectory in the $%
\left( V,\mu \right) $ plane was plotted to calculate $\partial \mu
/\partial V$, as per Eq.~(\ref{Bulkmodulus}).
\end{enumerate}
This procedure makes it possible to directly compare the VA predictions with
the numerically exact results.
\begin{figure}[tbp]
{\includegraphics[width=3.16in]{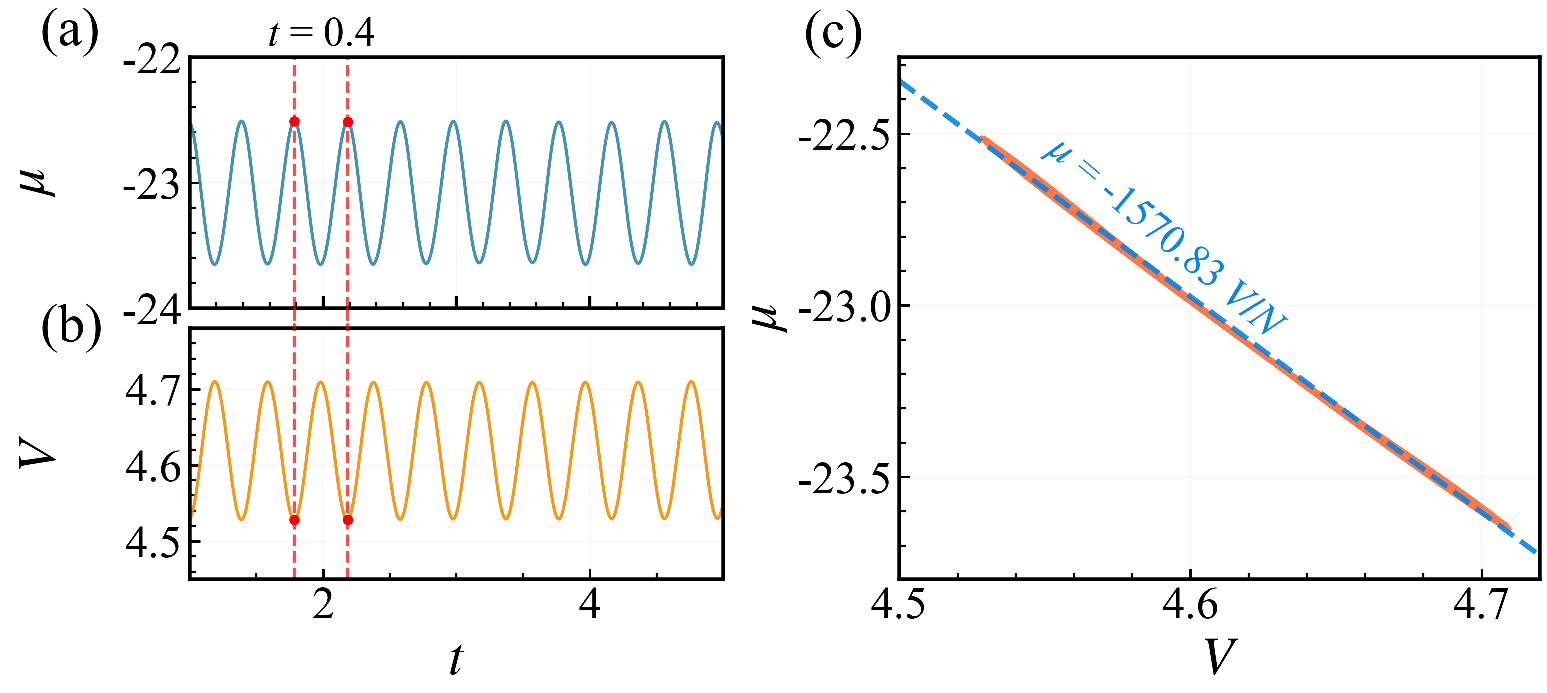}}
\caption{The quench dynamics of the QD with $\mathcal{N}=250$ and $g=-6$,
initiated by the weak perturbation, $g\rightarrow g+\protect\delta g$ with $%
\protect\delta g/g=0.01$. Panels (a,b) show the evolution of chemical
potential $\protect\mu $ and volume $V$, respectively, while panel (c)
depicts the corresponding trajectory in the $\left( V,\protect\mu \right) $
plane, which is actually a straight line. In (a,b), red dashed lines mark
two adjacent peaks of $\protect\mu $ and $V$, corresponding to an
oscillation period of $t=0.4$.}
\label{rtp-curve}
\end{figure}

For the case of $\mathcal{N}=250$ and $g=-6$,
Figs.~\ref{rtp-curve}(a,b) display the evolution of $\mu (t)$ and effective
volume $V(t)=\left( 4\pi /3\right) \bar{r}^{3}$, observed as the result of
the procedure. The red dashed vertical lines in the plots indicate two
successive extrema of the corresponding curves, from which the oscillation
period is determined as $t=0.4$, corresponding to the frequency $\Omega _{%
\mathrm{Num}}=15.7$, while, according to Eq.~(\ref{Omega2}), the VA predicts
the frequency $\Omega _{\mathrm{VA}}=16.3$, thus demonstrating sufficiently
high accuracy of the VA. Note that the maximum in panel (a) corresponds to
the minimum in (b), implying out-of-phase temporal oscillations of $\mu $
and $\Omega $.

The negative slope of the respective trajectory in the $\left( V,\mu \right)
$ plane in Fig. \ref{rtp-curve}(c) yields the BM value $B_{\mathrm{Num}%
}=1570.83$, according to Eq. (\ref{Bulkmodulus}), while the corresponding VA
prediction gives $B_{\mathrm{VA}}=1650.91$. Thus, the VA accuracy is
reliable for the prediction of the BM\ too.

\begin{figure}[tbp]
{\includegraphics[width=3.4in]{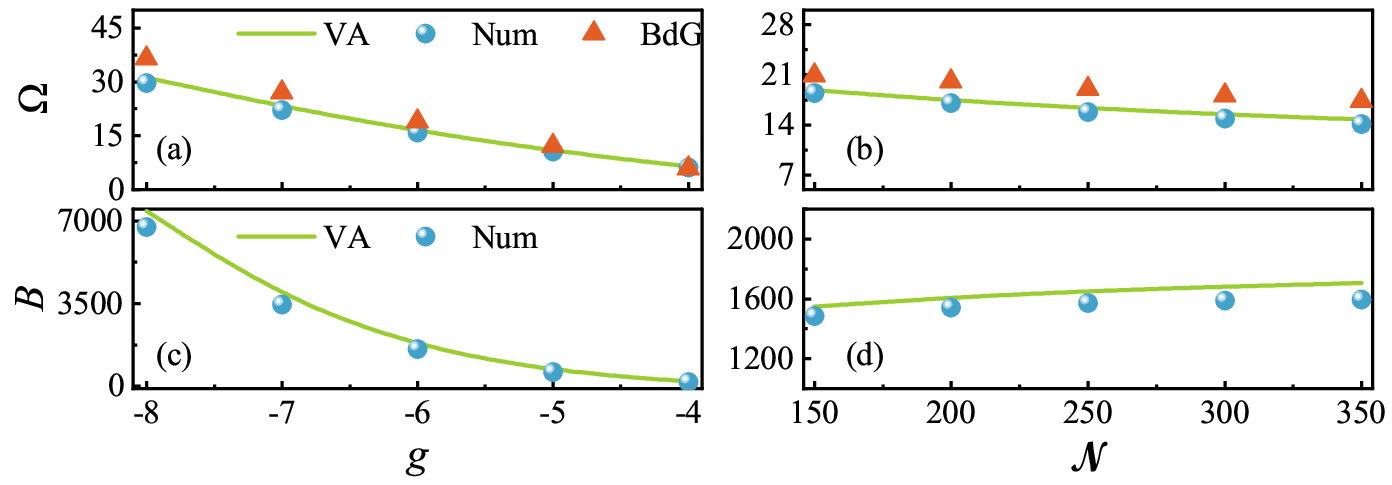}}
\caption{(a,b) and (c,d) The oscillation eigenfrequency $\Omega $ and BM $B$%
, respectively, vs. $g$ and $\mathcal{N}$. Green solid lines represent the
VA results, while chains of blue solid spheres denote the numerical results.
The red triangles in panels (a,b) correspond to the BdG calculations. In
panels (a,c), the norm is fixed as $\mathcal{N}=250$, whereas in panels
(b,d), the MF attraction strength is fixed as $g=-6$.}
\label{B-omega-gN}
\end{figure}

To summarize the results, we fix $\mathcal{N}=250$ and $g=-6$,
systematically comparing the numerical findings with the VA-predicted
oscillation eigenfrequency and BM, as shown in Figs. \ref{B-omega-gN}(a,b)
and (c,d). In Figs.~\ref{B-omega-gN}(a,b), the eigenfrequency $\Omega $ is
plotted vs. $g$ and $\mathcal{N}$ (naturally, for $g<0$, when the MF
nonlinearity is attractive, thus providing the existence of the self-trapped
QD states). The green solid line represents the VA predictions, given by
Eq.~(\ref{Omega2}), while their numerically found counterparts are shown by
chains of blue spheres.

The frequency of small intrinsic vibrations of the QDs can also be found as
the lowest nonzero real eigenfrequency from the numerical solution of the
Bogoliubov - de Gennes (BdG) equations for small perturbations of the wave
function, linearized around the stationary QD state. As seen in Figs.~\ref%
{B-omega-gN}(a,b), the VA predictions for $\Omega $ are in good agreement
with the numerical simulations and the BdG analysis alike.

For the BM, the VA and numerical results are again plotted, respectively, by
the green solid line and chain of blue solid spheres in Figs.~\ref%
{B-omega-gN}(c,d). It is seen that both stronger MF attraction (larger $-g$)
and larger norm $\mathcal{N}$ make the BM larger. These figures corroborate
the good agreement between the numerical and variational results.

\section{THE RELATION BETWEEN BULK MODULUS $B$ and frequency $\Omega$}
\label{sec:relation}
\begin{figure}[tbp]
{\includegraphics[width=3.4in]{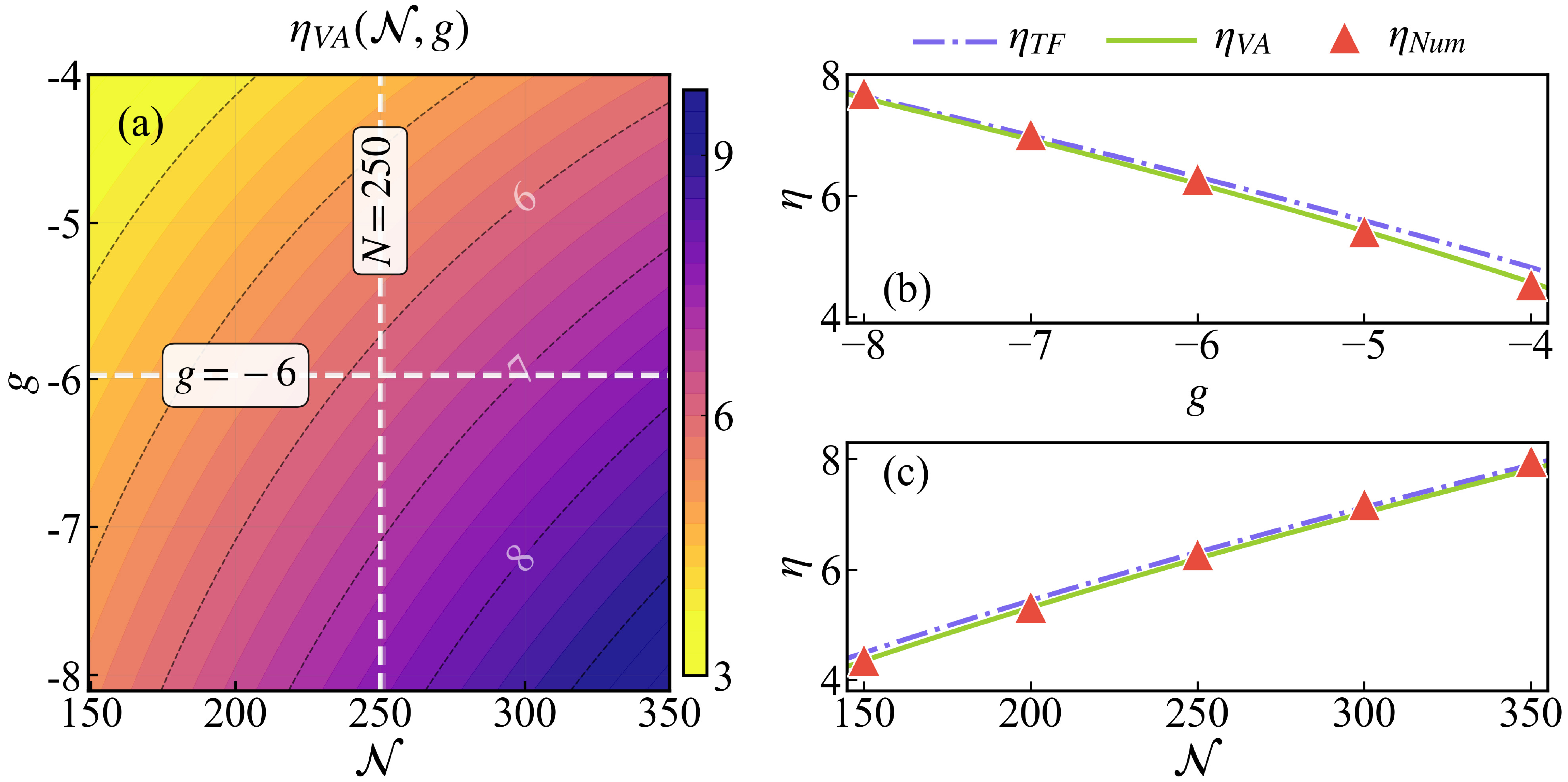}}
\caption{(a) The heatmap of values $\protect\eta_{\mathrm{VA}} (\mathcal{N},g)$ of the
BM/eigenfrequency ratio (\protect\ref{eta_formula}). The color shading from
light to dark indicates increasing values of $\protect\eta_{\mathrm{VA}} $, while the
black dashed curves correspond to the contour lines of $\protect\eta_{\mathrm{VA}} $. The
vertical and horizontal white dashed lines indicate $\mathcal{N}=250$ and $%
g=-6$, corresponding to the cases shown in panels (b) and (c), respectively.
Panels (b,c) display the dependence of $\protect\eta $ on $g$ and $\mathcal{N%
}$, where the green solid lines represent the VA result (\protect\ref%
{eta_relation}), the purple dashed-dotted line correspond to the TF
approximation given by Eq. (\protect\ref{eta-TF}), and the chain
of red triangles denotes the numerical results.}
\label{eta_N_g}
\end{figure}

In classical mechanics, the elastic BM and eigenfrequency of intrinsic
vibrations are subject to the well-known proportionality relation~\cite%
{SRaoB-Omega-relation},
\begin{equation}
B\propto \Omega ^{2}.  \label{class-B-Omega2}
\end{equation}%
Motivated by it, we seek to establish a similar relation between $B$ and $%
\Omega $ in the present setup. To this end, at the equilibrium point, we define
\begin{equation}
\eta = \left. \frac{B}{\Omega^2} \right|_{w = w_0}. \label{eta_formula}
\end{equation}%
The corresponding VA-predicted value $\eta _{\mathrm{VA}}$ can be obtained
from Eqs.~(\ref{Omega2}) and (\ref{VA-Bulkmodulus}), The result may be
presented in the form of
\begin{equation}
\eta _{\mathrm{VA}}=\kappa \frac{\mathcal{N}}{4\pi \bar{r}},
\label{eta_relation}
\end{equation}%
where $\kappa >0$ is a constant. As seen in Fig.~\ref{eta_N_g}%
(a), $\eta _{\mathrm{VA}}$ increases as a function of both the norm $%
\mathcal{N}$ and MF attraction strength $g$. Comparing Eq. (\ref%
{eta_relation}) with the numerical results across the $(\mathcal{N},g)$
parameter plane, we have found that the best agreement is achieved at
$\kappa =0.32$, for which the deviation of
$\eta _{\mathrm{VA}}$ from its numerically found counterpart remains below $5\%$
throughout the entire plane.
\begin{figure}[tbp]
{\includegraphics[width=2.8in]{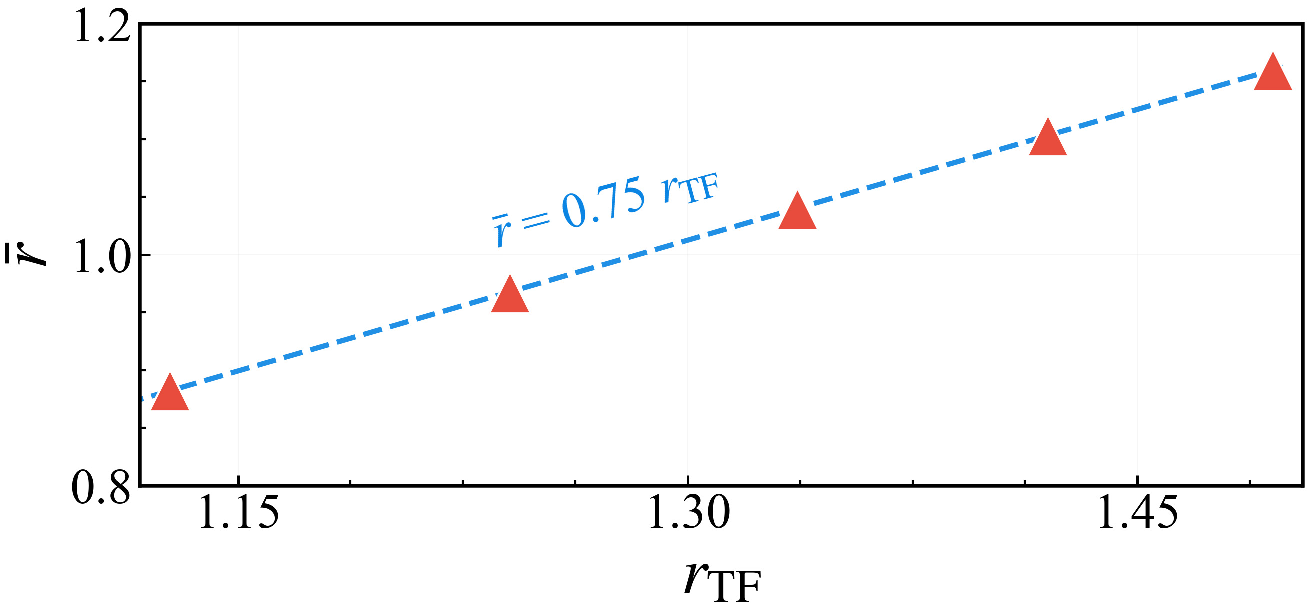}}
\caption{The relation between $r_{\mathrm{TF}}$ and $\bar{r}$ for $g=-6$ and
$\mathcal{N}=150\sim350$. The red triangles represent the numerically obtained
values of $\bar{r}$ sampled at $\mathcal{N}=150, 200, 250, 300, 350$, while
the blue dashed line shows the Thomas-Fermi prediction for the characteristic
radius, $r_{\mathrm{TF}} = \left( 27 \mathcal{N} / 25 \pi g^{2} \right)^{1/3}$.
}
\label{r-TF-vs-r}
\end{figure}

The Thomas-Fermi (TF) approximation, which yields a characteristic radius $%
r_{\mathrm{TF}}=\left( 27\mathcal{N}/25\pi g^{2}\right) ^{1/3}$ (see Appendix \ref{eta-TF-Sm}),
can also be applied in the
present context. Comparing the TF radius $r_{\mathrm{TF}}$ with
the $\bar{r}$, as shown in Fig. (\ref{r-TF-vs-r}),
we find that $\bar{r} \approx \tfrac{3}{4} r_{\mathrm{TF}}$.
Substituting this in Eq. (\ref{eta_relation}) yields the following relation:
\begin{equation}
\eta_{\mathrm{TF}} = \kappa \frac{\mathcal{N}}{3\pi r_{\mathrm{TF}}}
= \frac{\kappa}{9} \left( \frac{5g\mathcal{N}}{\pi} \right)^{2/3}.
\label{eta-TF}
\end{equation}

Figs.~\ref{eta_N_g}(b,c) present the comparison between the VA, TF
approximations and numerical resules for computing the BM/eigenfrequency ratio $\eta $, with
parameters fixed as $\mathcal{N}=250$ and $g=-6$. It is seen that the three
approaches are in full agreement. The
results for $\eta$ reveal a connection between
BM and eigenfrequency, which is governed by $\mathcal{N}$ and $g$. This connection makes it possible
to obtain the value of BM
from the measurement of the QD's internal vibrational frequency $\Omega$.

\section{Relationship between RMS-based and TF-based bulk modulus for flat-top droplets}
\label{meaning-r-bar}
As illustrated in Fig.~\ref{r-TF-vs-r}, the root-mean-square
(RMS) radius $\bar r$ scales linearly with the Thomas--Fermi (TF) radius $r_{\mathrm{TF}}$.
 For the QD in the large-$\mathcal{N}$ limit, the
density profile exhibits a flat-top shape. Assuming uniform density
$n_0$ for $r < r_{\text{TF}}$ and a vanishing density for $r > r_{\text{TF}}$,
the RMS radius is analytically determined as
\begin{equation}
    \begin{split}
        \bar{r} &= \sqrt{\frac{\int r^{2}|\psi|^{2}d^{3}\mathbf{r}}
        {\int |\psi|^{2}d^{3}\mathbf{r}}} = \sqrt{\frac{\int_{0}^{r_{\text{TF}}}
        r^2 \cdot n_0 \cdot 4\pi r^2 dr}{\int_{0}^{r_{\text{TF}}}
        n_0 \cdot 4\pi r^2 dr}} \\
        &= \sqrt{\frac{3}{5}}r_{\text{TF}} \approx 0.7746r_{\text{TF}}.
    \end{split}
\end{equation}
These considerations elucidate the relationship observed in Fig.~\ref{r-TF-vs-r},
which stems fundamentally from the distinct boundary assumptions and the corresponding
treatment of the density profile.
Furthermore, considering the effective volume $V$ defined in Eq.~(\ref{V}), we obtain
\begin{equation}
    V = \frac{4\pi}{3}\bar{r}^3 = \frac{4\pi}{3}(0.7746r_{\text{TF}})^3
    \approx 0.465 V_{\text{TF}},
\end{equation}
where $V_{\text{TF}} = \frac{4\pi}{3}r_{\text{TF}}^3$ represents the TF
volume of the droplet.
This scaling significantly affects the estimate of the bulk modulus $B$.
Specifically, the modulus is defined as
\begin{equation}
    B = -\mathcal{N} \frac{\partial \mu}{\partial V}
    = -\mathcal{N} \frac{\partial \mu}{\partial (0.465 V_{\text{TF}})}
    \approx 2.15 B_{\text{TF}},
\end{equation}
where $B_{\text{TF}}$ denotes the modulus derived for the TF volume.
Consequently, the bulk modulus defined via the RMS radius overestimates
the TF value of the flat-top QD by a factor of $\approx 2$.

In fact, the deviations arise from the different density profiles underlying the
two approaches: a sharp-cutoff profile in the TF approximation, and the VA-predcited smoothly
decaying profile. These differences lead to moderate quantitative
variations in the elastic modulus, while preserving the same overall behavior.

\section{experimental estimation}
\label{sec:experiment}

\begin{table*}
\caption{Correspondence between dimensionless and physical quantities.}\centering
\renewcommand{\arraystretch}{1.2}
 \begin{tabular*}{0.8\textwidth}{@{\extracolsep{\fill}} c c c}
\hline
\textbf{Dimensionless quantity} & \textbf{Scaling relation} & \textbf{%
Physical quantity} \\ \hline
$x,\,y,\,z = 1$ & $(x, y, z)\, l_{0}/\sqrt{\gamma} = (X, Y, Z)$ & $(X,Y,Z)\approx 1~\mathrm{\mu m}$ \\[4pt]
$t = 1$ & $t\, t_{0}/\gamma = t\, (M l_{0}^{2}) / (\hbar \gamma) = T$ & $T\approx0.7~\mathrm{ms}$ \\[4pt]
$\mathcal{N} = 1$ & $\mathcal{N}\, \gamma^{-3/2}= N$ & $N\approx~1.254\times10^3$ \\[4pt]
$g = -1$ & $g= 2\pi\delta a/(l_0\gamma)$ & $\delta a\approx-2.5a_{0}$ \\[4pt]
$B=1$ & $B\, (\hbar^2 \gamma)/(Ml_0^5)=\mathcal{B}$ & $\mathcal{B}\approx0.15~\mathrm{nPa}$ \\[4pt]
$\Omega=1$ & $\Omega\, (\hbar \gamma)/(Ml_0^2)=\omega$ & $\omega\approx1.4~\mathrm{kHz}$ \\ \hline
\end{tabular*}%
\label{tab:scaling}
\end{table*}


With this scaling, the correspondence between the dimensionless and physical
quantities is summarized in Table~\ref{tab:scaling}. For a droplet
characterized by $(g,\mathcal{N}) = (-6, 250)$, the corresponding total
atom number is
$3.13\times10^{5}$. The dimensionless values $\Omega \approx 16$ and $B\approx1600$ correspond to 
the actual physical quantities frequency $22.4~\mathrm{kHz}$ and the bulk modulus $0.24~\mu\mathrm{Pa}$.

\section{Conclusion}
\label{sec:conclusion}
In the framework of the three-dimensional GPE
(Gross-Pitaevskii equation), including the LHY (lee-Huang-Yang) correction
to the MF (mean-field) self-attraction of the binary condensate, we have
employed the VA (variational approximation) to derive analytical expressions
for the eigenfrequency of intrinsic vibrations and BM (bulk modulus) of
spatially isotropic QDs (quantum droplets), with numerical simulations
confirming the accuracy of the analytical results. The BM increases with the
atom number, while the eigenfrequency decreases, both growing with the
strength of the MF attraction. To reveal the relation between the BM and
eigenfrequency, we have introduced their ratio $\eta =B/\Omega ^{2}$ and,
combining the VA with the TF (Thomas-Fermi) approximation, we have obtained
the dependence of $\eta $ on the MF attraction strength and atom number. The
results suggest an experimental approach to determine the QDs' BM from the
measurement of the vibration frequency, and a possibility to explore and
employ the elasticity of quantum matter.

As an extension of the present analysis, it will be interesting to
investigate the elasticity of QDs in lower dimensions, where the LHY
correction takes a different form \cite{Petrov-low-dimensional}. A promising
direction is to explore the elasticity of QDs with more sophisticated
geometries, such as droplets carrying embedded vorticity \cite{2Dvort,3Dvort},
featuring a spatially modulated mean-field nonlinearity $g(r)$ \cite{Li-Elongated-PRA},
or governed by an anisotropic dipole-dipole interaction \cite{TPfau_nature,TPfau_review,ZYC_arbitrary_DDI}.
Another interesting extension is to investigate the elasticity of QDs in the strongly 
nonlinear regime of large-amplitude oscillations, where the density profile dynamically 
evolves between flat-top and Gaussian shapes.

\begin{acknowledgments}
We appreciate valuable discussions with Prof. Yiming Pan. This work was
supported by NNSFC (China) through Grants No. 12274077, No. 12475014,
Guangdong Basic and Applied Basic Research Foundation No. 2024A1515030131,
No. 2025A1515011128, No. 2023A1515110198, No. 2023A1515010770, the Research Fund of
Guangdong-Hong Kong-Macao Joint Laboratory for Intelligent Micro-Nano
Optoelectronic Technology through grant No. 2020B1212030010. 
\end{acknowledgments}

\appendix
\section{The derivation of the BM\ (bulk modulus)}

\label{derivation-B} In Ref. \cite{BEC-Ol}, for a system with
fixed particle number and in its ground state, the bulk modulus is defined as:
\begin{equation}
B=-V\frac{\partial p}{\partial V},
\end{equation}%
where $p$ is the pressure. For an approximately homogeneous density,
the pressure can be expressed as
\begin{equation}
\begin{split}
p=&-\frac{\partial E}{\partial V}=-\frac{\partial \left( \epsilon (n)V\right)
}{\partial V}=-\epsilon (n)-V\frac{\partial \epsilon (n)}{\partial V}\\
=&-\epsilon (n)+\frac{\mathcal{N}}{V}\frac{\partial \epsilon (n)}{\partial n},
\label{pressure}
\end{split}
\end{equation}%
where $\epsilon (n)$ is the energy density, $V$ is the effective volume, and
$n$ is the atom number density. Therefore, the bulk modulus can be written
as
\begin{equation}
\begin{split}
B=&-V\frac{\partial p}{\partial V}=-V\left( -\frac{\partial \epsilon (n)}{%
\partial V}+\frac{\partial \left( \frac{\mathcal{N}}{V}\frac{\partial
\epsilon (n)}{\partial n}\right) }{V}\right)\\ =&-V\frac{\mathcal{N}}{V}\frac{%
\partial \mu }{\partial V}=-\mathcal{N}\frac{\partial \mu }{\partial V},
\label{der-bulkmodulus}
\end{split}
\end{equation}%
which corresponds to Eq. (\ref{Bulkmodulus}) in the main text.




\section{The Thomas-Fermi (TF) approximation}
\label{eta-TF-Sm}
The density of the confined state with norm $\mathcal{N}$
and volume $V$ is $n=\mathcal{N}/V$, provided that the density is nearly
constant, see Fig. \ref{pk-omega-B}. The QD featuring a nearly flat-top
density profile, the energy functional Eq.~(\ref{energy}) can be simplified
under the TF approximation: as
\begin{equation}
\begin{split}
E=&\int \left( \frac{1}{2}gn^{2}+\frac{2}{5}n^{5/2}\right) d^{3}\mathbf{r}%
\\ &=\left( \frac{1}{2}gn^{2}+\frac{2}{5}n^{5/2}\right) V\\ &\equiv \frac{1}{2}%
\mathcal{N}gn+\frac{2}{5}\mathcal{N}n^{3/2}.  \label{TF-energy}
\end{split}
\end{equation}%
The equilibrium condition for the density is
\begin{equation}
\frac{dE}{dn}=\frac{1}{2}\mathcal{N}g+\frac{3}{5}\mathcal{N}n^{1/2}=0,
\label{dE-dn-0}
\end{equation}%
which yields the equilibrium value of the density,
\begin{equation}
n_{\mathrm{e}}=\frac{25}{36}g^{2},  \label{equilibrium-n}
\end{equation}%
the corresponding equilibrium value of the volume being
\begin{equation}
V_{\mathrm{e}}=\frac{\mathcal{N}}{n_{\mathrm{e}}}=\frac{36}{25}\frac{%
\mathcal{N}}{g^{2}}.
\end{equation}%
Using the definition of the effective volume as per Eq.~(\ref{V}), the
corresponding QD's radius is
\begin{equation}
r_{\mathrm{TF}}=\left( \frac{3V}{4\pi }\right) ^{1/3}=\left( \frac{27%
\mathcal{N}}{25\pi g^{2}}\right) ^{1/3}.  \label{r-TF}
\end{equation}%
By substituting it for $\bar{r}$ in Eq.~(\ref{eta_relation}), one finally
obtains Eq. (\ref{eta-TF}) in the main text.
Substituting $\mathcal{N}$ and $g$ into the above equation (\ref{r-TF}) and comparing with $\bar{r}$
under identical parameters, we find that $\bar{r} \approx \tfrac{3}{4} r_{\mathrm{TF}}$,
as illustrated in the Fig. \ref{r-TF-vs-r}.

\end{document}